\documentclass[prb,reprint,a4paper,citeautoscript,floatfix]{revtex4-2}

\pdfoutput=1
\usepackage[charter]{mathdesign}
\usepackage{amsmath}

\usepackage[pdftex]{graphicx}
\usepackage{microtype}
\usepackage{bm}\let\vec\bm

\usepackage[svgnames]{xcolor}
\usepackage[
	colorlinks=True,linkcolor=DarkRed,citecolor=ForestGreen,urlcolor=MediumBlue,
	pdfstartview=FitH,bookmarks=False,pdfpagemode=UseNone
]{hyperref}

\def\kB{k_{\mathrm{B}}}
\def\RH{R_{\mathrm{H}}}
\def\rr{R_{\mathrm{H}}^{}/R_{\mathrm{H}}^0}

\begin{document}

\title{Hall response of locally correlated two-dimensional electrons at low density}

\author{Giacomo Morpurgo}
\affiliation{Department of Quantum Matter Physics, University of Geneva, 1211 Geneva, Switzerland}
\author{Louk Rademaker}
\affiliation{Department of Quantum Matter Physics, University of Geneva, 1211 Geneva, Switzerland}
\author{Christophe Berthod}
\affiliation{Department of Quantum Matter Physics, University of Geneva, 1211 Geneva, Switzerland}
\author{Thierry Giamarchi}
\affiliation{Department of Quantum Matter Physics, University of Geneva, 1211 Geneva, Switzerland}

\date{January 29, 2024}

\begin{abstract}

We study the Hall constant in a homogeneous two-dimensional fluid of correlated electrons immersed in a perpendicular magnetic field, with special focus on the regime of low carrier density. The model consists of a one-band tight-binding model and a momentum-independent causal self-energy, representing interaction-induced correlations effects that are restricted to be local in space. We write general gauge-invariant equations for the conductivity tensor at first order in the magnetic field and solve them numerically---analytically when possible---in a minimal model of anisotropic square lattice with constant self-energy. Our results show that deviations from the universal behavior of the Hall constant, as observed in the semiclassical regime, appear upon entering the quantum regime, where the Fermi energy and interaction are comparable energy scales.

\end{abstract}

\maketitle

\section{Introduction}

In the wake of graphene's discovery \cite{Novoselov-2004}, the study of gated two-dimensional (2D) semiconductors has developed swiftly, providing a whole family of new low-density conductors. The achievable carrier concentrations are sufficiently low---as low as $10^{10}$--$10^{11}$~cm$^{-2}$---to even realize the elusive interaction-dominated Wigner crystal \cite{Smolenski-2021, Zhou-2021, Li-2021}. In gated moir\'{e} heterostructures, recent experiments have provided a wealth of transport data in the low-density regime, even through density-controlled metal-insulator transitions \cite{Mak-2022}. In recent years, 2D magnetic insulators and conductors have been uncovered as well \cite{Gibertini-2019, Wang-2022}, some of which like CrSBr display puzzling magnetotransport properties \cite{Wu-2022}. This situation prompts further theoretical research on the magnetotransport in 2D correlated conductors at low carrier densities. 

Doing so is not easy, since one needs to properly describe assemblies of charge carriers in a regime where Fermi energy, scattering rate, and temperature are comparable energy scales. This means that some of the usual approximations such as the Boltzmannian semiclassical theory \cite{Ziman-1960} cannot be applied. This theory indeed regards carriers as true particles characterized by well-defined energy and momentum and subject to Fermi statistics. However, this semiclassical description reaches its limits when the quasiparticle scattering rate is not small compared to either the quasiparticle energy---i.e., typically $\kB T$, where $T$ is the temperature---or the effective Fermi energy, which is proportional to the carrier density. In this case, the spectral function of the electrons is broad and one must use the quantum theory \cite{Fukuyama-1969-1, Itoh-1985}. However, the general first-principles many-body approach poses considerable computational challenges and progress has been difficult. 

For high densities, the Hall response of correlated electrons was studied in the Hubbard model using exact numerical methods on small systems \cite{Assaad-1995, Wang-2020} or approximate methods \cite{Pruschke-1995, Markov-2019, Vucicevic-2021}, both at weak \cite{Assaad-1995, Pruschke-1995, Wang-2020} and strong magnetic field \cite{Markov-2019, Vucicevic-2021}. Progress could be made for interacting systems in special situations such as quasi-one-dimensional systems, made of weakly coupled, strongly interacting chains \cite{Leon-2006, Leon-2007, Leon-2008-1} or, more recently, for ladders for which the Hall effect can be accessed both numerically and analytically \cite{Filippone-2019, Greschner-2019-1, Buser-2021} and probed in cold atomic gases experiments \cite{Zhou-2023}.

Extending such methods to the case of vanishing low densities would be highly non trivial, so we follow here another route: we neglect the nonlocal correlations between the charge carriers. At the single-particle level, this means that the electrons are characterized by a dispersion relation $E_{\vec{k}}$ and a complex self-energy function $\Sigma(\varepsilon)$ that is local (independent of ${\vec{k}}$), like in the dynamical mean-field theory \cite{Georges-1992, *Georges-1996}. The absence of vertex corrections, both at zero \cite{Khurana-1990} and finite magnetic field \cite{Markov-2019, Vucicevic-2021}, simplifies the calculation of the conductivities significantly. This class of models thus provides an interesting playground, intermediate between the approximate semiclassical theory and the intractable quantum theory, where the magnetotransport can be studied accurately in the thermodynamic limit. In this approach, the self-energy may be regarded as a phenomenological input or taken from a microscopic calculation.

In the present paper, we use this local self-energy approach to study the Hall effect in systems with low densities. The plan of the paper is as follows. In Sec.~\ref{sec:theory} we present general gauge-invariant transport equations that capture the dc Hall response of a 2D single-band tight-binding model of locally correlated electrons. Given a dispersion relation and a causal self-energy, the transport equations can be solved numerically in all regimes of density and temperature. In Sec.~\ref{sec:results}, we solve these equations in a minimal model, where the correlations are represented by a single scattering rate $\Gamma$. The simplicity of the model allows us to obtain exact asymptotic results, in particular, in the low-density regime. We compare the resulting Hall constant $\RH$ with the universal value $\RH^0=-1/(n|e|)$. Quite remarkably, we find that in the model with a self-energy having neither momentum nor energy dependence, the Hall effect at low density is enhanced to four times the semiclassical result. Moreover, we find a non monotonic evolution of the Hall factor with a pronounced maximum when the temperature is raised to values larger than the scattering rate. This effect is most pronounced at low density, large scattering rate, and for a strongly anisotropic (quasi-one-dimensional) Fermi surface. We discuss further our findings in Sec.~\ref{Sec:Conclusion}. Some of the more technical points can be found in the Appendices.

\section{System studied and transport equations}
\label{sec:theory}

\subsection{Model Hamiltonian and Green's function}

\begin{figure}[tb]
\includegraphics[width=\columnwidth]{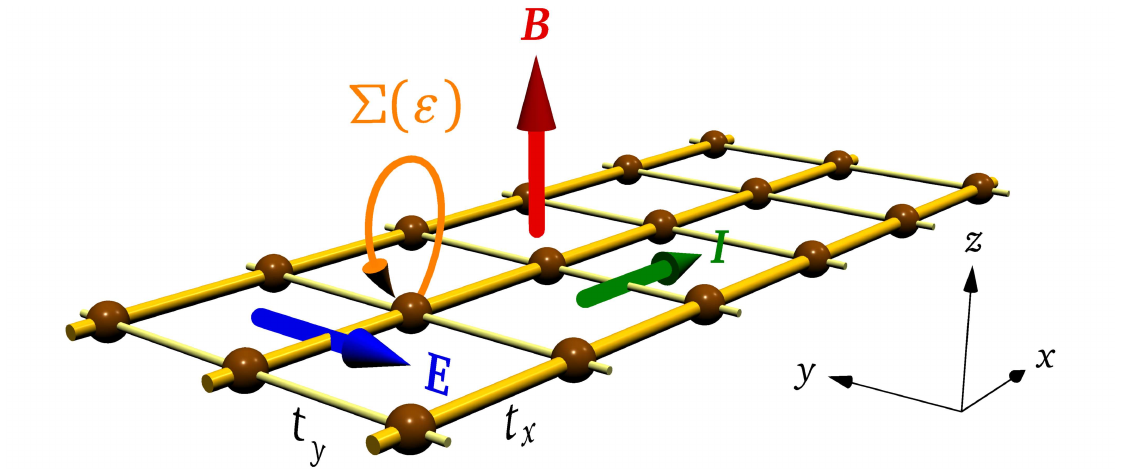}
\caption{\label{fig:model}
We consider a two-dimensional tight-binding model with direction-dependent hopping amplitudes $t_x$, $t_y$ and a magnetic field $B$ along the $z$ direction. We pass a current $I$ along the $x$ direction and measure the created electric field (Hall effect) along the direction $y$. Furthermore, we make the assumptions that interactions in the system can only lead to local correlations and thus a self-energy $\Sigma(\varepsilon)$ depending only on the energy $\varepsilon$. 
}
\end{figure}

We study a 2D interacting one-band tight-binding model without spin-orbit coupling immersed in a perpendicular magnetic field $\vec{B}$ as shown in Fig.~\ref{fig:model}. Its Hamiltonian is
    \begin{equation}\label{eq:H}
        H=\sum_{\vec{r}_1\vec{r}_2\sigma}t^0_{\vec{r}_1\vec{r}_2}
        e^{i\frac{|e|}{\hbar}\int_{\vec{r}_1}^{\vec{r}_2}d\vec{r}\cdot\vec{A}(\vec{r})}
        c^\dagger_{\vec{r}_1\sigma}c^{\phantom{\dagger}}_{\vec{r}_2\sigma} + H_{\mathrm{int}},
	\end{equation}
where $t^0_{\vec{r}_1\vec{r}_2}\equiv t^0_{\vec{r}_1-\vec{r}_2}$ is the translation-invariant zero-field hopping amplitude from site $\vec{r}_1$ to site $\vec{r}_2$ of the lattice, $c^\dagger_{\vec{r}\sigma}$, $c^{\phantom{\dagger}}_{\vec{r}\sigma}$ are the creation and annihilation operators at site $\vec{r}$ for spin $\sigma$, and $\vec{A}$ is the vector potential, which enters via the Peierls substitution. This Hamiltonian leads to a dispersion relation $E_{\vec{k}}=\sum_{\vec{\rho}}t^0_{\vec{\rho}}e^{-i\vec{k}\cdot\vec{\rho}}$ at $B=0$, where $\vec{k}$ is a wave vector and $\vec{\rho}$ is a vector linking two sites in the lattice. Although the model can be totally generic, we specialize in the applications that follow to a 2D anisotropic square lattice, which means that $t^0_{\vec{\rho}}$ will become, for nearest-neighbors $t_x$ and $t_y$, depending on the direction of the coupling. This anisotropy defines the smallest hopping as one additional energy scale that can be tuned to be comparable with either Fermi energy, scattering rate, or temperature. In this way, we scan a class of geometries ranging from the quasi-1D case to the 2D case (isotropic hoppings).

Moreover, we assume throughout this paper that the interactions between particles, described by a Hamiltonian $H_{\mathrm{int}}$, lead only to local correlations and therefore to a \emph{local self-energy} in zero magnetic field. The self-energy $\Sigma(\varepsilon)$ thus depends only on the energy $\varepsilon$ and not on the direction and size of the momentum. As a consequence, the direction of the group velocity is not modified by interactions, which in the diagrammatic language can be seen as the absence of vertex corrections.

Using this self-energy allows us to define the zero-field spectral function $A(\vec{k},\varepsilon)=(-1/\pi)\mathrm{Im}\,G(\vec{k},\varepsilon)$, where the single particle Green's function is $G(\vec{k},\varepsilon)=[\varepsilon-E_{\vec{k}}-\Sigma(\vec{k},\varepsilon)]^{-1}$. For a local self-energy $\Sigma(\varepsilon)$, the spectral function only depends on $\vec{k}$ through $E_{\vec{k}}$ and it is convenient to define $A(\vec{k},\varepsilon)\equiv A(E_{\vec{k}},\varepsilon)$ with
	\begin{equation}\label{eq:A}
   		A(E,\varepsilon)=\frac{-\mathrm{Im}\,\Sigma(\varepsilon)/\pi}
    	{[\varepsilon-E-\mathrm{Re}\,\Sigma(\varepsilon)]^2+[\mathrm{Im}\,\Sigma(\varepsilon)]^2}.
	\end{equation}
Thus, in our notations, the variable $E$ refers to an energy shell of the noninteracting dispersion, while the variable $\varepsilon$ refers to the energy axis, over which the electron spectral weight is redistributed by the correlations encoded in $\Sigma(\varepsilon)$.

Note that our model is characterized only by the dispersion relation $E_{\vec{k}}$, which is linked to the geometry of the system, and the self-energy $\Sigma(\varepsilon)$, which encodes possible interactions and disorder. As a consequence, the conductivity will also be a function of only $E_{\vec{k}}$ and $\Sigma(\varepsilon)$. Note also that Eq.~(\ref{eq:A}) is not applicable in a \emph{finite} field, where the self-energy acquires an essential momentum dependence associated with the fractal Hofstadter spectrum \cite{Hofstadter-1976}. As we discuss in Appendix~\ref{app:CRJJ}, since the Hofstadter spectrum is even in $B$, Eq.~(\ref{eq:A}) is sufficient to evaluate perturbatively any property up to first order in $B$.

\subsection{Self-consistent chemical potential}

In the low-density limit that we study, and especially when the Fermi energy (i.e., some measure of the particle or hole density) is comparable with temperature and/or scattering rate, the precise knowledge of the chemical potential is crucial. We will compute it self-consistently using the relation
	\begin{equation}\label{eq:n}
		n=\int_{-\infty}^{\infty}d\varepsilon\,f(\varepsilon-\mu)
		\int_{-\infty}^{\infty}dE\,N_0(E)A(E,\varepsilon),
	\end{equation}
where $f(\varepsilon)=(e^{\varepsilon/\kB T}+1)^{-1}$ is the Fermi-Dirac distribution, $A(E,\varepsilon)$ is the spectral function defined previously in Eq.~(\ref{eq:A}), and $N_0(E)$ is the noninteracting density of states (DOS), which for a 2D band of spin-1/2 electrons is given by
	\begin{equation}\label{eq:N0}
		N_0(E)=2\int\frac{d^2k}{(2\pi)^2}\,\delta\left(E-E_{\vec{k}}\right).
	\end{equation}
The quantity $N(\varepsilon)=\int_{-\infty}^{\infty}dE\,N_0(E)A(E,\varepsilon)$ in Eq.~(\ref{eq:n}) is the interacting DOS. $N(\varepsilon)$ is, in general, nonzero outside the noninteracting band. As a result, the chemical potential may move below the noninteracting band at low electron density (above it at low hole density), even at zero temperature.

\subsection{Hall constant}

The Hall effect is the appearance of an electric field in the direction transverse to the current, which defines a transverse resistivity according to $E_y=\rho_{yx}J_x$. In the presence of nontrivial topology or broken time-reversal symmetry, an anomalous Hall effect occurs in zero magnetic field \cite{Nagaosa-2010}. The models considered in this paper are space- and time-inversion symmetric and have no spin-orbit coupling, and are therefore characterized by an ordinary Hall response, i.e., a resistivity $\rho_{yx}(B)$ proportional to $B$. The Hall constant is then defined as $\RH=\lim_{B\to0}\rho_{yx}(B)/B$. The transverse resistivity is related to the conductivity via $\rho_{yx}=\sigma_{xy}/(\sigma_{xx}\sigma_{yy}+\sigma_{xy}^2)$, where the property $\sigma_{yx}=-\sigma_{xy}$ has been used \footnote{The property $\sigma_{yx}=-\sigma_{xy}$ follows from space-inversion symmetry, which changes $J_x$ into $-J_x$ and $J_y$ into $-J_y$, and time-reversal symmetry, which changes $B$ into $-B$.}. An ordinary Hall response also implies that the correction to $\sigma_{xx}$ and $\sigma_{yy}$ is quadratic in $B$. Hence, the conductivity tensor has the structure
	\begin{equation}\label{eq:sigma}
		\sigma_{\alpha\beta}(B)=\delta_{\alpha\beta}\sigma^{(0)}_{\alpha}
		+(1-\delta_{\alpha\beta})\sigma^{(1)}_{\alpha\beta}B+O(B^2).
	\end{equation}
Using this, we obtain a representation of the Hall constant in terms of the expansion coefficients $\sigma^{(0)}_{\alpha}$ and $\sigma^{(1)}_{xy}$:
	\begin{equation}\label{eq:RH}
    	\RH=\lim_{B\to 0}\frac{1}{B}\frac{\sigma_{xy}}{\sigma_{xx}\sigma_{yy}+\sigma_{xy}^2}
    	=\frac{\sigma^{(1)}_{xy}}{\sigma^{(0)}_{x}\sigma^{(0)}_{y}}.
	\end{equation}
Following Ohm's law, $\sigma_{\alpha\beta}$ gives the linear response to the electric field and must therefore be calculated at vanishing electric field. In the next section, we first write expressions for $\sigma_{\alpha\beta}$ that contain all orders in the magnetic field $B$ using the Kubo formula. We then perform an expansion in $B$ to extract $\sigma^{(0)}_{\alpha}$ and $\sigma^{(1)}_{xy}$.

\subsection{dc limit of the Kubo conductivity tensor}

The macroscopic (i.e., $q=0$ or spatially averaged) complex ac conductivity at frequency $\omega$ is given by the Kubo formula \cite{Mahan-2000}:
	\begin{equation}\label{eq:Kubo}
		\sigma_{\alpha\beta}(\omega)=\frac{i}{\omega}\frac{1}{V}\left[
		C^R_{J_{\alpha}J_{\beta}}(\omega)-C^R_{J_{\alpha}J_{\beta}}(0)\right].
	\end{equation}
In this expression, $J_{\alpha}$ is the macroscopic current operator in the direction $\alpha$ (given in Appendix \ref{app:CRJJ}) and $C^R_{J_{\alpha}J_{\beta}}(\omega)$ is the retarded current-current correlation function:
	\begin{equation}\label{eq:CJJ}
		C^R_{J_{\alpha}J_{\beta}}(\omega)=\int_{-\infty}^{\infty}dt\,e^{i\omega t}(-i/\hbar)\theta(t)
		\langle[J_{\alpha}(t),J_{\beta}(0)]\rangle.
	\end{equation}
This is an extensive quantity that is divided by the volume $V$---the surface, in our 2D case---to yield an intensive conductivity. The second term in Eq.~(\ref{eq:Kubo}) is the diamagnetic response, which is real in normal metals and ensures that the dc conductivity is finite.

Our target is the dc conductivity that is purely real and relates to the imaginary part of the correlation function:
	\begin{equation}\label{eq:DC}
		\sigma_{\alpha\beta}=\lim_{\omega\to 0}\mathrm{Re}\,\sigma_{\alpha\beta}(\omega)
		=-\lim_{\omega\to 0}\frac{1}{\omega}\mathrm{Im}\,\frac{1}{V}
		C^R_{J_{\alpha}J_{\beta}}(\omega).
	\end{equation}
To evaluate $\sigma_{\alpha\beta}$, we express the currents in terms of the operators $c^\dagger_{\vec{r}\sigma}$ and $c_{\vec{r}\sigma}$ and, thanks to the absence of vertex corrections, recast the correlation function as an expression involving only the interacting single-particle Green's function. The latter is neither gauge- nor translation invariant in the presence of a magnetic field. Nevertheless, the correlation function is brought to a manifestly gauge-invariant form after introducing a modified gauge- and translation-invariant Green's function. The details are given in Appendix~\ref{app:CRJJ}.

The resulting expression of the current-current correlation function involves a transport function $\Phi_{\alpha\beta}$ that depends on $E_{\vec{k}}$ but not on $\Sigma(\varepsilon)$, analogous to the DOS in Eq.~(\ref{eq:N0}), and the spectral function Eq.~(\ref{eq:A}) that depends on $\Sigma(\varepsilon)$ but not on $E_{\vec{k}}$:
    \begin{multline}\label{eq:CRJJ}
		C^R_{J_{\alpha}J_{\beta}}(\omega)=V\int_{-\infty}^{\infty}d\varepsilon_1d\varepsilon_2\,
		\frac{f(\varepsilon_1-\mu)-f(\varepsilon_2-\mu)}{\hbar\omega+\varepsilon_1-\varepsilon_2+i0}\\
		\times\int_{-\infty}^{\infty}dE_1dE_2\,\Phi_{\alpha\beta}(E_1,E_2,B)
		A(E_1,\varepsilon_1)A(E_2,\varepsilon_2).
	\end{multline}
The number of energy integrals is doubled with respect to Eq.~(\ref{eq:n}), reflecting the occurrence of two current operators. This is also the case for the transport function, which now involves two momentum sums:
	\begin{multline}\label{eq:Phiab}
		\hspace{-1em}
		\Phi_{\alpha\beta}(E_1,E_2,B)=-\left(\frac{e}{\hbar}\right)^2\frac{1}{V}\sum_{\vec{k}\vec{q}\sigma}
		\delta\left(E_1-E_{\vec{k}-\frac{\vec{q}}{2}}\right)
		\delta\left(E_2-E_{\vec{k}+\frac{\vec{q}}{2}}\right)\\
		\times\sum_{\vec{\rho}_1\vec{\rho}_2}
		\rho_{1\alpha}t^0_{\vec{\rho}_1}e^{i(\vec{k}+\vec{q}/2)\cdot\vec{\rho}_1}
		\rho_{2\beta}t^0_{\vec{\rho}_2}e^{i(\vec{k}-\vec{q}/2)\cdot\vec{\rho}_2}\\
		\times\frac{1}{N}\sum_{\vec{r}}e^{i\vec{q}\cdot\vec{r}}
		e^{\frac{i|e|}{2\hbar}\vec{B}\cdot\vec{r}\times(\vec{\rho}_1-\vec{\rho}_2)}.
	\end{multline}
Although the transport function contains all orders in $B$, Eq.~(\ref{eq:CRJJ}) is not valid to all orders in $B$, as it misses the Landau-level physics. On a lattice, this breaks the spectral function into a non-perturbative fractal Hofstadter spectrum \cite{Hofstadter-1976}. It is nevertheless possible to show that the variation of the spectral function is quadratic in $B$ (Appendix~\ref{app:CRJJ}). Therefore, Eqs.~(\ref{eq:CRJJ}) and (\ref{eq:Phiab}) are valid at first order in $B$.

The dc limit of the conductivity involves the imaginary part of $C^R_{J_{\alpha}J_{\beta}}(\omega)$, which lets two terms appear. The first combines $-\pi\delta(\hbar\omega+\varepsilon_1-\varepsilon_2)$ from the imaginary part of the retarded energy denominator in Eq.~(\ref{eq:CRJJ}) with the real part of the transport function. This term denoted $\sigma_{\alpha\beta}^{(e)}(B)$ is even in $B$ and reads
	\begin{multline}\label{eq:sigmae}
		\sigma_{\alpha\beta}^{(e)}(B)=\pi\hbar\int_{-\infty}^{\infty}d\varepsilon\,[-f'(\varepsilon-\mu)]\\
		\times\int_{-\infty}^{\infty}dE_1dE_2\,\mathrm{Re}\,\Phi_{\alpha\beta}(E_1,E_2,B)
		A(E_1,\varepsilon)A(E_2,\varepsilon),
	\end{multline}
where $f'$ denotes the energy derivative of $f$. It is a Fermi-surface or on-the-energy-shell contribution, due to the factor $-f'(\varepsilon-\mu)$ that constrains the energy $\varepsilon$ to remain within a few $\kB T$ around the chemical potential.

The second term involves the principal value of the energy denominator and the imaginary part of the transport function. Upon expanding $1/(\hbar\omega+\varepsilon_1-\varepsilon_2)$ for $\omega\to0$, a contribution proportional to $1/(\varepsilon_1-\varepsilon_2)$ appears. This part is identically zero by symmetry, due to the fact that the imaginary part of the transport function is odd under the exchange of $E_1$ and $E_2$. The subsequent term of the expansion is $-\hbar\omega/(\varepsilon_1-\varepsilon_2)^2$, which once inserted in Eq.~(\ref{eq:DC}) yields a well-behaved limit that is odd in $B$ and is denoted $\sigma_{\alpha\beta}^{(o)}(B)$:
	\begin{multline}\label{eq:sigmao}
		\sigma_{\alpha\beta}^{(o)}(B)=\hbar\int_{-\infty}^{\infty}d\varepsilon_1d\varepsilon_2\,
		\frac{f(\varepsilon_1-\mu)-f(\varepsilon_2-\mu)}{(\varepsilon_1-\varepsilon_2)^2}\\
		\times\int_{-\infty}^{\infty}dE_1dE_2\,\mathrm{Im}\,\Phi_{\alpha\beta}(E_1,E_2,B)
		A(E_1,\varepsilon_1)A(E_2,\varepsilon_2).
	\end{multline}
Unlike $\sigma_{\alpha\beta}^{(e)}(B)$, this appears to be an off-the-energy-shell term with the energies $\varepsilon_1$ and $\varepsilon_2$ allowed to take values far away from the chemical potential.

\subsection{Expansion for weak field}

The transport function in Eq.~(\ref{eq:Phiab}) contains all the magnetic-field dependence, but it is cumbersome for numerical evaluation. We therefore consider its expansion in powers of $B$. The zero-field limit is easy to grasp: the $\vec{r}$ sum forces $\vec{q}$ to vanish, which in turn forces $E_1$ and $E_2$ to be equal; the remaining $\vec{\rho}_1$ and $\vec{\rho}_2$ sums yield derivatives of the dispersion $E_{\vec{k}}$. As a result, the transport function is real and Eq.~(\ref{eq:sigmae}) yields the first coefficient appearing in Eq.~(\ref{eq:sigma}):
	\begin{align}
    	\label{eq:sigma0}
    	\sigma_{\alpha}^{(0)}&=\pi\hbar\int_{-\infty}^{\infty}d\varepsilon\,[-f'(\varepsilon-\mu)]
    	\int_{-\infty}^{\infty}dE\,\Phi_{\alpha}^{(0)}(E)A^2(E,\varepsilon)\\
    	\label{eq:Phi0}
		\Phi_{\alpha}^{(0)}(E)&=\left(\frac{e}{\hbar}\right)^22\int\frac{d^2k}{(2\pi)^2}
		\left(\frac{\partial E_{\vec{k}}}{\partial k_{\alpha}}\right)^2\delta(E-E_{\vec{k}}).
	\end{align}
Details are provided in Appendix~\ref{app:expansion}. The interpretation of this formula is that the dc conductivity is controlled by zero-energy electronic transitions close to the chemical potential. On every shell of energy $E$, the transitions are weighted by the transport function, which is proportional to the squared $\alpha$ component of the group velocity averaged on that shell, and by the square of the spectral function, which sets the amount of electronic spectral weight available for the initial and final states of the transition.

At first order in $B$, the transport function is imaginary and we only get a contribution from Eq.~(\ref{eq:sigmao}). Remarkably, we find that the off-shell character of this term is not fundamental, as the expression can be recast \emph{exactly} in an on-shell form (Appendix~\ref{app:expansion}):
	\begin{align}
		\label{eq:sigma1}
		\sigma_{xy}^{(1)}&=\hbar\int_{-\infty}^{\infty}d\varepsilon\,[-f'(\varepsilon-\mu)]
		\int_{-\infty}^{\infty}dE\,\Phi^{(1)}(E)A^3(E,\varepsilon)\\
		\nonumber
		\Phi^{(1)}(E)&=\left(\frac{|e|}{\hbar}\right)^3\frac{2\pi^2}{3}\int\frac{d^2k}{(2\pi)^2}
		\left[2\frac{\partial E_{\vec{k}}}{\partial k_x}
		\frac{\partial E_{\vec{k}}}{\partial k_y}
		\frac{\partial^2E_{\vec{k}}}{\partial k_x\partial k_y}\right.\\
		\label{eq:Phi1}
		&\quad\left.-\left(\frac{\partial E_{\vec{k}}}{\partial k_x}\right)^2
		\frac{\partial^2E_{\vec{k}}}{\partial k_y^2}
		-\left(\frac{\partial E_{\vec{k}}}{\partial k_y}\right)^2
		\frac{\partial^2E_{\vec{k}}}{\partial k_x^2}\right]\delta(E-E_{\vec{k}}).
	\end{align}
The transformation from off-shell form to on-shell form relies explicitly on the causal character of the spectral function, which in turns requires the causality of the self-energy. The latter must therefore respect the Kramers--Kronig relations. We have not been able to bring the full Eq.~(\ref{eq:sigmao}) to on-shell form, a frustration that has already been expressed in the literature \cite{Itoh-1985}.

Let us make some comments on our results and, in particular, the expression for $\sigma_{xy}^{(1)}$, in connection with previous works. Our derivation recovers some known results that are scattered across the relevant literature, but proceeds differently from previous works. In Refs.~\cite{Fukuyama-1969-1, Itoh-1985, Voruganti-1992, Mitscherling-2018, Pickem-2022}, the limit of vanishing magnetic field is taken from the outset, before the dc limit, which allows the authors to relate the Hall response with a three-current correlation function. Within the family of locally correlated models, the dc limit can be taken before expanding in powers of the magnetic field. The resulting expression for the transverse conductivity is off-shell, as found by other authors \cite{Itoh-1985, Markov-2019, Vucicevic-2021}. As we showed, at first order in the magnetic field, this apparently off-shell contribution can be rigorously recast, after the expansion in $B$, as a on-shell expression involving the \emph{third power} \cite{Voruganti-1992, Pruschke-1995, Mitscherling-2018, Pickem-2022} of the spectral function in Eq.~(\ref{eq:sigma1}), which is reminiscent of the three-current correlation function that characterizes the transverse response when the low-field expansion is performed before taking the dc limit \cite{Fukuyama-1969-1, Itoh-1985, Voruganti-1992}. This equivalence relies explicitly on the causality of the self-energy function.

Second, the average between the $x$ and $y$ directions appearing at the second line of Eq.~(\ref{eq:Phi1}) reflects the gauge invariance of the formula and is crucial in our anisotropic setting. Related expressions based on the Boltzmann theory in the literature lack this symmetrization and should therefore be corrected for anisotropic systems \cite{Ziman-1960, Fukuyama-1969-2, Ong-1991}. In Appendix~\ref{app:Boltzmann}, we show how to recover from Eqs.~(\ref{eq:sigma0})--(\ref{eq:Phi1}) the Boltzmann theory in the isotropic relaxation-time approximation.

\section{Results for the anisotropic square lattice}
\label{sec:results}

Results are presented here for the Hall constant of the anisotropic tight-binding model depicted in Fig.~\ref{fig:model}, with a special emphasis on the regime of low electron density. For the self-energy, we use the simplest causal model, $\Sigma(\varepsilon)=-i\Gamma$, which represents quasiparticles with an energy-independent relaxation time $\tau=\hbar/(2\Gamma)$. While elementary, this model has two advantages. It contains a single energy scale to describe the scattering, like the semiclassical theory in the isotropic relaxation-time approximation, such that differences between the quantum and semiclassical approaches cannot be ascribed to differences in the scattering. Furthermore, it allows for a better numerical control, as certain integrals can be performed exactly (see Appendix~\ref{app:convolutions} and Ref.~\cite{Pickem-2023}). This same model of scattering was used in recent studies of the Hall response in multiband systems \cite{Mitscherling-2020, Pickem-2021, Mitscherling-2022, Pickem-2022}.

One peculiarity of this self-energy is that the spectral function---and \textit{a fortiori} the interacting DOS---have $\sim1/\varepsilon^2$ tails without low- or high-energy cutoff. Consequently, the chemical potential is also unbounded in the limit of vanishing density, even at zero temperature. Microscopic lattice models generally have self-energies with a finite support, possibly different from the support of the noninteracting band. The low-density behavior of $\RH$ may depend on the details of the self-energy in nontrivial ways, and additional work is needed to determine how the results presented here would change if a cutoff were introduced in the self-energy.

\subsection{DOS and transport functions}

\begin{figure}[tb]
\includegraphics[width=0.8\columnwidth]{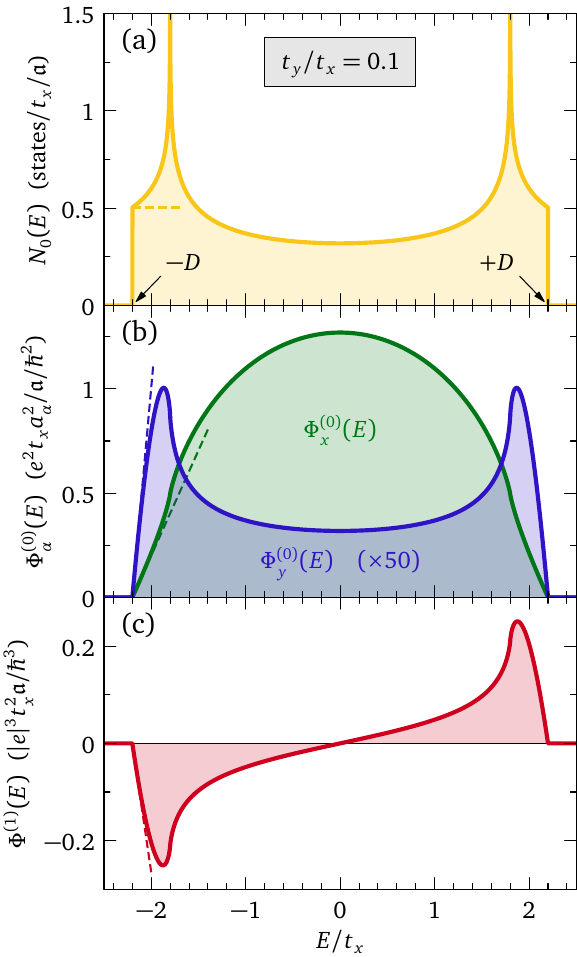}
\caption{\label{fig:N0-Phi}
(a) Density of states, (b) zeroth-order transport functions, and (c) first-order transport function for the model shown in Fig.~\ref{fig:model} with $t_y/t_x=0.1$. Energies are measured in units of $t_x$ and $\mathfrak{a}=a_xa_y$ is the area of the unit cell. The dashed lines show the leading behavior close to the band edge $E=-D$, i.e., $N_0(E)=1/(2\pi\mathfrak{a}\sqrt{|t_xt_y|})$, $\Phi^{(0)}_{\alpha}(E)=(e/\hbar)^2(a_{\alpha}^2/\mathfrak{a})|t_{\alpha}|/\sqrt{|t_xt_y|}\frac{1}{\pi}(E+D)$, and $\Phi^{(1)}(E)=-(|e|/\hbar)^3\mathfrak{a}\sqrt{|t_xt_y|}\frac{4\pi}{3}(E+D)$.
}
\end{figure}

Before discussing the conductivities, we briefly describe the DOS and the transport functions of the system sketched in Fig.~\ref{fig:model}. The DOS is displayed in Fig.~\ref{fig:N0-Phi}(a) for $t_y/t_x=0.1$. It is particle-hole symmetric with a half-bandwidth $D=2(|t_x|+|t_y|)$. There are two logarithmic van Hove singularities at energies $\pm2(|t_x|-|t_y|)$, that merge into one in the isotropic 2D case and become two square-root singularities at the band edges in the 1D case. The graph of $\Phi_{\alpha}^{(0)}(E)$ is displayed in Fig.~\ref{fig:N0-Phi}(b). Like the DOS, it is particle-hole symmetric. Where the DOS has van Hove singularities, it displays logarithmic singularities of the first derivative. Figure~\ref{fig:N0-Phi}(c) shows $\Phi^{(1)}(E)$, that is odd for the model of Fig.~\ref{fig:model}, which is consistent with the fact that the Hall constant must vanish at half filling in particle-hole symmetric systems. Interestingly, we find that $\Phi^{(1)}(\mu)$ is equal to $(|e|/\hbar)^3(4\Gamma^2/3)A_l$, where $A_l$ is the Stokes area introduced by Ong \cite{Ong-1991}, if the latter is calculated for an isotropic scattering time $\tau_{\vec{k}}=\hbar/(2\Gamma)$.

\subsection{Zero-temperature Hall constant}

\begin{figure}[tb]
\includegraphics[width=\columnwidth]{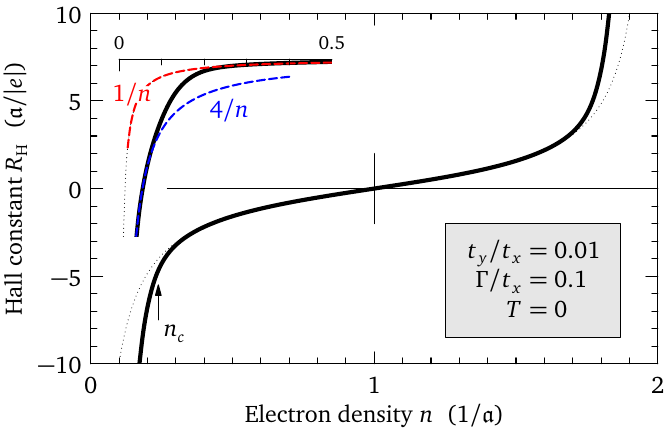}
\caption{\label{fig:RH-n}
Zero-temperature Hall constant in the anisotropic square lattice with $t_y=0.01t_x$ and $\Gamma=0.1t_x$. Dotted lines show Eq.~(\ref{eq:RH1}) and $n_c$ indicates Eq.~(\ref{eq:nc}). The inset highlights the behavior at low electron density. $\mathfrak{a}$ is the area of the unit cell.
}
\end{figure}

The strength of nonclassical effects may be quantified by comparing the Hall constant $\RH$ with the universal value $\RH^0=-1/(n|e|)$. Even in the semiclassical regime, band-structure effects are known to reduce the Hall factor $r=\rr$ to values smaller than unity \cite{Ong-1991}. For nearly free electrons, Ref.~\cite{Fukuyama-1969-1} found that the momentum dependence of the self-energy renormalizes the Hall constant by a factor $r=(N_0/N)^2$, with $N_0$ and $N$ the bare and renormalized Fermi-level densities of states, which is also generally smaller than unity. Another finding of Ref.~\cite{Fukuyama-1969-1} was that the energy dependence of the self-energy yields no renormalization of the Hall constant.

For our model, Fig.~\ref{fig:RH-n} shows a typical evolution of the Hall constant with varying carrier density at $T=0$, for parameters $t_y/t_x=0.01$ and $\Gamma/t_x=0.1$. $\RH$ is negative (positive) for electron (hole) conduction and diverges at low electron or hole density. It vanishes at $n=1$ due to the particle-hole symmetry of the model. At low electron/hole density, the divergences seen are expected from the semiclassical theory, which predicts $-1/n$, respectively, $1/(2-n)$ behavior. However, a closer inspection reveals that the divergence is four times faster than the semiclassical prediction (see inset of Fig.~\ref{fig:RH-n}). A Hall factor $r$ \emph{higher} than unity is normally not expected \cite{Fukuyama-1969-1, Ong-1991}. Here, the value $r=4$ occurs in the quantum regime, where the semiclassical assumption of sharp quasiparticles is no longer justified.

To better understand the low-density asymptotic behaviors in these two regimes, we consider first the semiclassical case by taking the limit $\Gamma\to 0^+$ in the transport equations at $T=0$ (see Appendix~\ref{app:asymptoticsT=0}). We find the Hall constant
	\begin{equation}\label{eq:RH1}
		\RH=\frac{3}{2\pi^2\hbar}\frac{\Phi^{(1)}(\mu)}{\Phi^{(0)}_x(\mu)\Phi^{(0)}_y(\mu)}
		\quad(T=0, \Gamma\to0).
	\end{equation}
This result simplifies even further in the limit of low density: As $n$ approaches zero, the chemical potential approaches the lower band edge, where the DOS becomes constant and the transport functions vanish linearly (see Fig.~\ref{fig:N0-Phi}).  We can therefore substitute in Eq.~(\ref{eq:RH1}) the limiting expressions given in the legend of Fig.~\ref{fig:N0-Phi} and use the limiting value of the density $n=N_0(-D)(\mu+D)$, which yields the semiclassical result $\RH^0=-1/(|e|n)$. Away from the band edge, $\RH$ as given by Eq.~(\ref{eq:RH1}) is smaller than $\RH^0$,  irrespective of the anisotropy $t_y/t_x$. This is consistent with the expectation that band-structure effects reduce the Hall constant in the semiclassical approximation \cite{Ong-1991}. The full semiclassical result of Eq.~(\ref{eq:RH1}) is displayed in Fig.~\ref{fig:RH-n} as dotted lines, to be compared with the full quantum result obtained using Eqs.~(\ref{eq:RH}) and (\ref{eq:sigma0})--(\ref{eq:Phi1}).

Instead, in the quantum regime, one must keep $\Gamma$ finite while taking the limit $n\to0$. Due to the $\sim1/\varepsilon^2$ tails of the interacting DOS, this implies sending $\mu$ to $-\infty$. We find that $\mu\propto-\Gamma/n$ as $n\to0$ (see Appendix~\ref{app:asymptoticsT=0}). At the same time, the normal and Hall conductivities behave as $\sigma_{\alpha}^{(0)}\sim n^4/(4\Gamma^2)$ and $\sigma_{xy}^{(1)}\sim -n^7/(4\Gamma^4)$, such that we indeed find
	\begin{equation}\label{eq:RH4}
		\RH=-\frac{4}{|e|n}\quad(T=0, n\to0).
	\end{equation}
It is remarkable that, although the conductivities have completely different forms in the quantum and semiclassical regimes (in the latter, they read $\sigma_{\alpha}^{(0)}\sim n/\Gamma$ and $\sigma_{xy}^{(1)}\sim -n/\Gamma^2$), the Hall constant is nevertheless independent of the scattering and proportional to $1/n$ in both cases.

\begin{figure}[tb]
\includegraphics[width=\columnwidth]{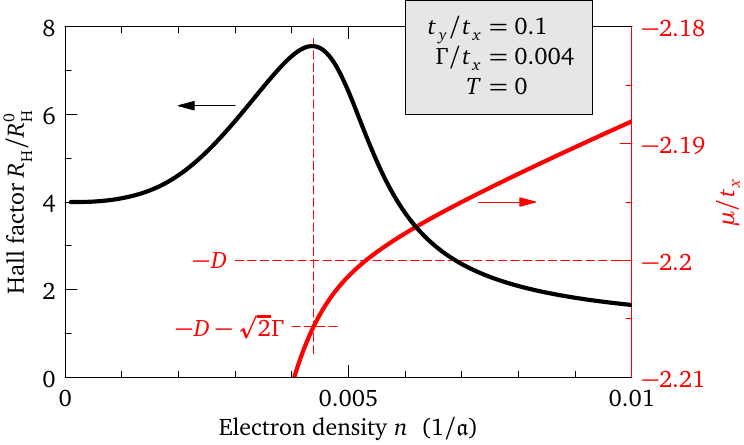}
\caption{\label{fig:RH-mu}
Hall factor $r=\rr$ (left axis) and chemical potential (right axis) versus electron density at $T=0$ for $t_y/t_x=0.1$ and $\Gamma/t_x=0.004$.
}
\end{figure}

\begin{figure*}[tb]
\includegraphics[width=\textwidth]{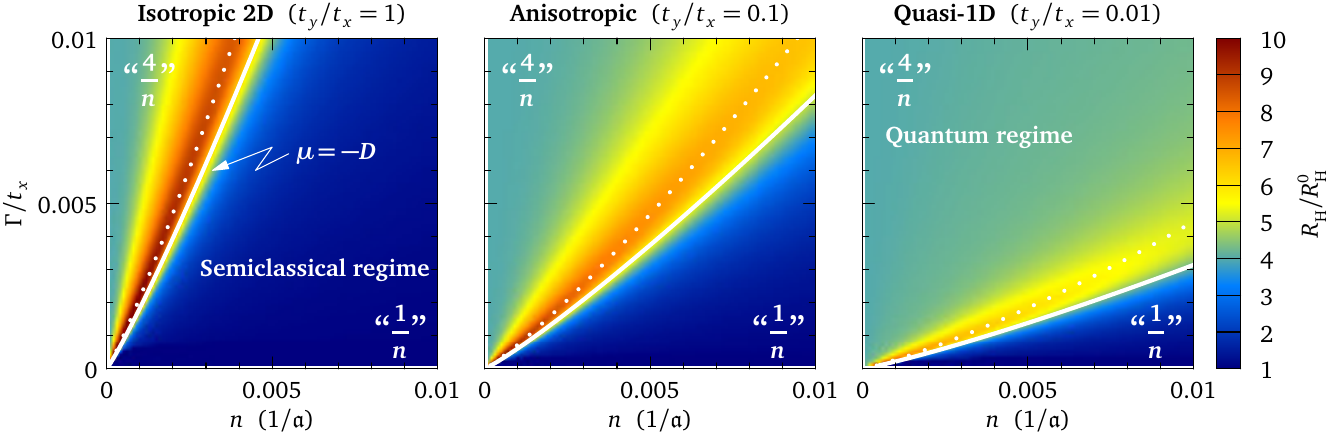}
\caption{\label{fig:RH-T=0}
Hall factor $r=\rr$ at $T=0$ and low electron density for three values of the anisotropy $t_y/t_x$. Dark blue corresponds to $r=1$ and turquoise to $r=4$. The semiclassical and quantum regimes are labeled with $1/n$ and $4/n$, respectively. The solid (dotted) white lines correspond to $\mu=-D$ ($\mu=-D-\sqrt{2}\Gamma$) and mark the beginning (approximate maximum) of the crossover between the semiclassical and quantum regimes.
}
\end{figure*}

With the aim to determine the characteristic density $n_c$ at which $\RH$ crosses over from the semiclassical regime to the quantum regime, we plot in Fig.~\ref{fig:RH-mu} the density dependence of the Hall factor, superimposed with the density dependence of the chemical potential. The Hall factor varies non monotonously between $r=1$ and $r=4$ as $n$ decreases, going through a maximum when $\mu$ lies approximately $\sqrt{2}\Gamma$ below the band edge. Hence, the semiclassical to quantum crossover is associated with the chemical potential approaching and crossing the bottom of the noninteracting band. From the condition $\mu=-D$, we deduce an approximate expression for $n_c$ (see Appendix~\ref{app:asymptoticsT=0}):
	\begin{equation}\label{eq:nc}
		n_c=\frac{\Gamma}{2\pi^2\mathfrak{a}\sqrt{|t_xt_y|}}\left[1+\ln\left(\frac{2D}{\Gamma}\right)\right].
	\end{equation}
Hence, the quantum regime is reached at higher densities for larger scattering rates and more one-dimensional geometries. These trends are visualized in Fig.~\ref{fig:RH-T=0}, where the zero-temperature Hall factor is displayed versus $n$, $\Gamma$, and $t_y/t_x$. Upon going from isotropic 2D to quasi-1D, $n_c$ increases and the quantum regime progressively expands. Around $n_c$, the density dependence of the conductivities $\sigma_{xx}$ and $\sigma_{yy}$ crosses over from $\sim n^4$ to $n$ and the Hall conductivity from $n^7$ to $n$. The steeper increase of $\sigma_{xy}$ for $n\lesssim n_c$ explains the appearance of a maximum in the $n$-dependence of the Hall factor. We shall not describe this mechanism further here, as a similar one occurs as a function of temperature and is discussed at length below.

\subsection{Temperature dependence}

Since at zero temperature the strongest deviation from a semiclassical Hall effect occurs when the chemical potential lies slightly outside the noninteracting band, we also expect an enhancement of the Hall factor when $\mu$ moves in the vicinity of the band edge under the effect of temperature. As we will see, this enhancement can reach several orders of magnitude in our model. In models where the scattering rate has a cutoff, a finite temperature will also drive $\mu$ below the band if the density is sufficiently low, such that similar effects could also possibly occur. Note that inelastic scattering mechanisms usually lead to $T$-dependent self-energies. By keeping the scattering rate $\Gamma$ independent of $T$, we ensure that the $T$-dependent effects we find are not due to changes in the scattering. The study of $T$-dependent scattering models is left for future works.

\begin{figure}[b]
\includegraphics[width=\columnwidth]{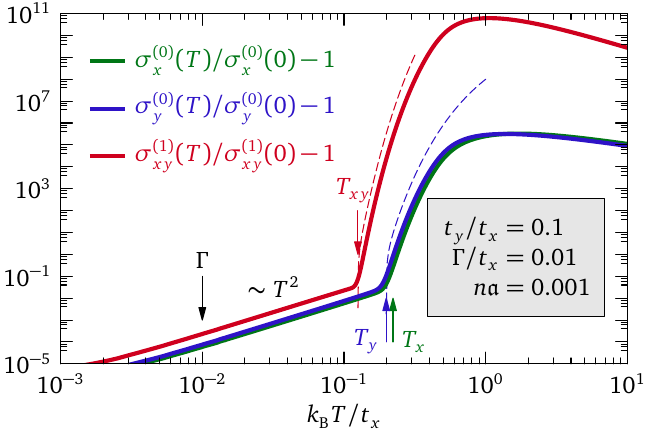}
\caption{\label{fig:T2}
Typical temperature dependence of the conductivities in the quantum regime ($t_y=0.1t_x$, $\Gamma=0.01t_x$, $n=0.001/\mathfrak{a}$). The dashed lines show Eq.~(\ref{eq:sigma-above-onset}). The onset temperatures indicated by arrows correspond to Eq.~(\ref{eq:onsets}).
}
\end{figure}

Starting at near-zero temperature, the conductivities and the Hall constant vary initially slowly like $T^2$. In Fermi liquids, the $T^2$ raise of the resistivity is due to the scattering rate growing like $(\varepsilon-\mu)^2$. Here, since we assume an energy- and temperature-independent self-energy, the $T^2$ dependence of the conductivities stems from thermal excitations rather than from the scattering rate. It occurs both in the semiclassical and quantum regimes. However, whether the conductivities increase or decrease like $T^2$ depends on the density (see Appendix~\ref{app:asymptoticsT>0}). In the quantum regime $n\to0$, we find that all conductivities \emph{increase} with a coefficient proportional to $(n/\Gamma)^2$. In the semiclassical regime, the coefficient depends on the slope and curvature of the transport function and can have different signs in different directions. Figure~\ref{fig:T2} shows the typical temperature variation of the conductivities in the quantum regime. While $\sigma_x^{(0)}$ and $\sigma_y^{(0)}$ have nearly the same $T^2$ coefficient, the coefficient of $\sigma^{(1)}_{xy}$ is larger, consistent with the asymptotic results given in Appendix~\ref{app:asymptoticsT>0}.

Strikingly, the conductivities shoot up by orders of magnitude at well-defined onset temperatures that are different for the $x$, $y$, and $xy$ components. Below the onset, the Fermi tails in Eqs.~(\ref{eq:sigma0}) and (\ref{eq:sigma1}) do not overlap significantly with the band, but only with the band \emph{tails} resulting from the convolution of the transport function with the second or third power of the spectral function. Physically, the thermally excited carriers are very few, occupying the tail of the spectral function. Above the onset, the Fermi tails overlap with the band and the number of thermally excited carriers increases strongly. As the figure shows, this occurs in a regime where $\kB T>\Gamma$, such that the spectral function is narrower than the derivative of the Fermi function. Using this fact, and linearizing the transport functions close to the band edge, as appropriate at low density, we derive in Appendix~\ref{app:asymptoticsT>0} approximations for the conductivities in this regime:
	\begin{subequations}\label{eq:sigma-above-onset}\begin{align}
    	\sigma_{\alpha}^{(0)}&\approx
		\frac{e^2}{h}\frac{a_{\alpha}^2}{\mathfrak{a}}\frac{|t_{\alpha}|}{\sqrt{|t_xt_y|}}
		\frac{\kB T}{\Gamma}\ln\left(1+e^{\frac{\mu+D}{\kB T}}\right)\\
		\sigma^{(1)}_{xy}&\approx-\frac{|e|^3}{h^2}2\pi\mathfrak{a}\frac{\sqrt{|t_xt_y|}\kB T}{\Gamma^2}
		\ln\left(1+e^{\frac{\mu+D}{\kB T}}\right).
	\end{align}\end{subequations}
To estimate the onset temperatures, we ignore the weak $T^2$ variation below the onset and determine the temperature at which the above expressions equal the zero-temperature conductivities. The resulting equations are simplified using the fact that $\exp\left(\frac{\mu+D}{\kB T}\right)\ll 1$, allowing an explicit solution:
	\begin{subequations}\label{eq:onsets}\begin{align}
		\kB T_{\alpha}&=\frac{\delta-D}
		{W\left(\frac{\delta(\delta-D)}{|t_{\alpha}|(|t_xt_y|)^{1/2}(\pi n\mathfrak{a})^3}\right)}\\
		\kB T_{xy}&=\frac{\delta-D}
		{W\left(\frac{\delta^2(\delta-D)}{2(|t_xt_y|)^{3/2}(\pi n\mathfrak{a})^5}\right)}.
	\end{align}\end{subequations}
Here $W(\cdot)$ is the Lambert function and we have defined $\delta=2\Gamma/(\pi n\mathfrak{a})$. These temperatures are displayed in Fig.~\ref{fig:T2}.

Figure~\ref{fig:T2} shows that $T_{xy}<T_x, T_y$, which can be rationalized as follows. According to Eq.~(\ref{eq:sigma-above-onset}), the conductivities behave as $\sigma_{x,y}\sim I(T)/\Gamma$ and $\sigma_{xy}\sim I(T)/\Gamma^2$ above onset, with the same temperature dependence $I(T)$ but different powers of $\Gamma$. The $T$ dependence follows from convolving the Fermi factor with the linearized transport function: $\int_{-D}^{\infty}d\varepsilon\,[-f'(\varepsilon-\mu)](\varepsilon+D)=\kB T\ln\left[1+\exp\left(\frac{\mu+D}{\kB T}\right)\right]$. The difference stems from the sum rule for $A^2(E,\varepsilon)$ being $\sim1/\Gamma$, while the sum rule for $A^3(E,\varepsilon)$ is $\sim1/\Gamma^2$, which reflects the fact that $A^3(E,\varepsilon)$ is narrower than $A^2(E,\varepsilon)$. Theses sum rules appear after convolving the  transport functions with $A^2$ and $A^3$ in Eqs.~(\ref{eq:sigma0}) and (\ref{eq:sigma1}), respectively. On the other hand, we have seen that, at $T=0$ and in the quantum regime, the conductivities behave as $\sigma_{x,y}\sim n^4/\Gamma^2$ and $\sigma_{xy}\sim n^7/\Gamma^4$. The equation for $T_{x,y}$ is therefore of the form $I(T)/\Gamma=n^4/\Gamma^2$, while the equation for $T_{xy}$ is of the form $I(T)/\Gamma^2=n^7/\Gamma^4$. Since $I(T)$ increases monotonically with increasing $T$, one sees that $T_{xy}<T_x,T_y$, provided that $n^3<\Gamma$. With all numeric and dimensional factors in place, this condition reads $(\pi n\mathfrak{a})^3<\Gamma/|t_x|$. In the regime $\kB T_{xy}>\Gamma$, where Eq.~(\ref{eq:onsets}) was obtained, this condition is met if $n<n_c$. We therefore expect that the relation $T_{xy}<T_x,T_y$ is robust throughout the quantum regime.

\begin{figure}[tb]
\includegraphics[width=\columnwidth]{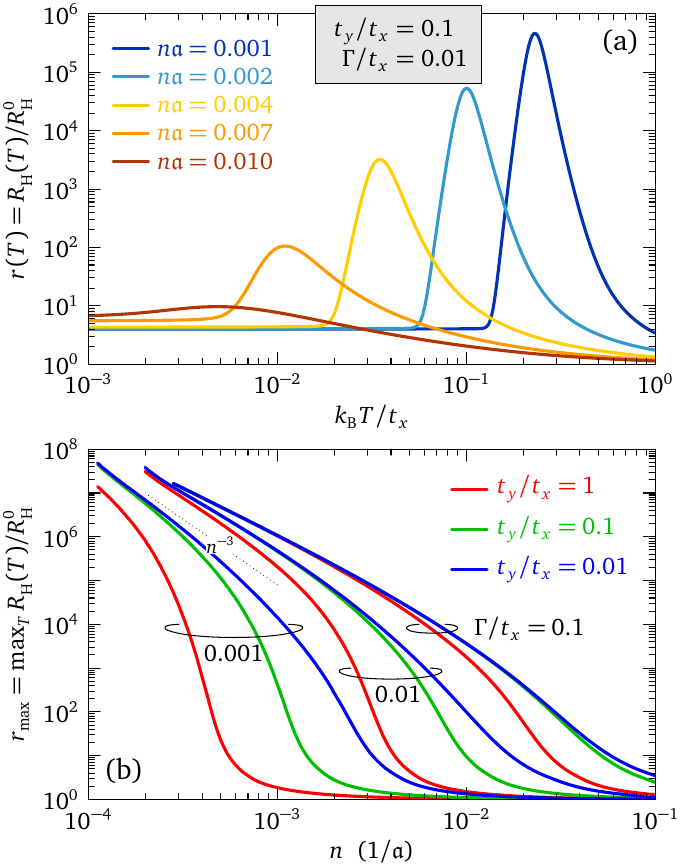}
\caption{\label{fig:RH-T}
(a) Temperature dependence of the Hall factor $r(T)=\RH^{}(T)/\RH^0$ for $t_y=0.1t_x$, $\Gamma=0.01t_x$, and various densities. (b) Largest value of the temperature-dependent Hall factor versus density for three values of the anisotropy $t_y/t_x$ and three values of the scattering rate $\Gamma$.
}
\end{figure}

At temperatures between $T_{xy}$ and $T_y$, the sharp increase of $\sigma_{xy}$ drives a correspondingly large increase of $\RH$, since the normal conductivities remain low. One expects a peak in $\RH(T)$ that reaches its maximum close to $T_y$. Evaluating Eq.~(\ref{eq:sigma-above-onset}) at $T=T_y$, we obtain the following estimate for the maximum Hall factor:
	\begin{equation}\label{eq:rmax}
		r_{\max}\approx\frac{4\Gamma}{|t_y|(\pi n\mathfrak{a})^3}.
	\end{equation}
This can be huge in the low-density limit, especially for a quasi-1D dispersion. Figure~\ref{fig:RH-T}(a) displays $\RH(T)$ in the anisotropic case $t_y/t_x=0.1$ with $\Gamma/t_x=0.01$ and at various densities that scan from deep in the quantum regime up to the crossover into the semiclassical regime (see Fig.~\ref{fig:RH-T=0}). In the quantum regime, $r(T)$ starts at the value $4$, raises slowly as $T^2$ (Appendix~\ref{app:asymptoticsT>0}), develops a strong maximum, and finally drops to the high-temperature value $r(\kB T\gg\Gamma)\approx 1$ (Appendix~\ref{app:asymptoticsT>0}). Consistently with Eq.~(\ref{eq:rmax}), the height of the maximum is very large at small $n$ and drops rapidly as $n$ increases. Upon approaching the semiclassical regime, the initial value $r(0)$ grows (Fig.~\ref{fig:RH-T=0}) and the maximum disappears. Finally, deep in the semiclassical regime (not shown), $r(T)$ varies with almost no temperature dependence from a value close to $1$ at $T=0$ to a value close to $1$ at high temperature.

Figure~\ref{fig:RH-T}(b) shows the maximum Hall factor determined numerically versus $n$, for three values of $t_y$ and three values of $\Gamma$. The power law $\sim n^{-3}$ predicted by Eq.~(\ref{eq:rmax}) is observed at low $n$, while for larger $n$ the behavior changes. The increase of $r_{\max}$ with increasing $\Gamma$ and decreasing $t_y$ is also qualitatively consistent with Eq.~(\ref{eq:rmax}).

\section{Summary and Conclusion}
\label{Sec:Conclusion}

Measuring the Hall constant remains the number-one experimental approach for determining the carrier density of conductors, including correlated ones (see for instance Ref.~\cite{Badoux-2016}). This is grounded in the universality of this property, as predicted by the Boltzmann semiclassical theory of transport. The conditions under which this universality breaks down are not well-known. The semiclassical approach relies on the carriers being true particles with fermionic statistics, which is likely a good approximation when interactions are weak and carrier density is sufficiently high. Deviations may be expected at low density, even before the regime of Wigner crystallization is reached, when the Fermi energy is comparable with the scattering rate, because the carriers cannot be treated as sharp quasiparticles. In this case, the semiclassical theory gives way to the Kubo--Greenwood framework, but a complete solution is generally out of reach. One exception, in which a closed set of equations may be written for the Hall constant, is the class of models, where nonlocal correlations between the electrons are either absent, like in dynamical mean-field theory, or negligible. We have collected these equations in Sec.~\ref{sec:theory}, for the case of a 2D single-band system in perpendicular magnetic field [Eqs.~(\ref{eq:A})--(\ref{eq:RH}), and (\ref{eq:sigma0})--(\ref{eq:Phi1})]. Our goal is to explore the predictions of these equations in the regime of low density, which is relevant in the present context of flourishing research activity on gated 2D conductors.

As a first application, we have studied a minimal two-parameters model. The first parameter, $t_y/t_x$, controls the electronic dimensionality and allows one to study 2D to 1D crossovers. The second, $\Gamma/t_x$, controls the scattering. We have studied $r\equiv\rr$ as a function of density and temperature and found deviations from the semiclassical regime characterized by $r=1$. As a rule of thumb, the semiclassical behavior is realized at higher dimensionality, weak scattering, high density, and high temperature, while significant deviations are found at low dimensionality, strong scattering, low density, and low temperature. There are two kinds of deviations from universality. At zero temperature, when the chemical potential moves outside the noninteracting band, the Hall constant crosses over from $1/n$ to $4/n$. This quantitative change does not completely break the universality, in the sense that the Hall constant remains asymptotically independent of band structure and scattering rate. However, the crossover region is nonuniversal, depending on both band structure and scattering rate [see Eq.~(\ref{eq:nc})]. A second kind of deviation arises at finite temperature, when thermally excited carriers---that were at lower temperatures confined to interaction-induced tails of the DOS---begin to occupy the extreme levels of the noninteracting band. The ensuing boost of conductivity, that occurs first for the Hall conductivity, produces a peak in the Hall factor $r$.

This model is peculiar in that the scattering rate has no dependence on the energy, which allows charge carriers to exist at arbitrary energies below the free-carrier band. In a way, it may be seen as the simplest generalization of the Boltzmann theory in the isotropic relaxation time approximation: It reduces to the latter in the semiclassical regime, and allows one to describe the crossover into the quantum regime without additional parameters. This circumstance enables good analytical and numerical control of the equations as well as exact asymptotic expressions but may also lead to peculiar results. Nevertheless, the model has the merit of showing that significant deviations from the universal Hall constant paradigm can occur at low density and gives a first hint on where to search the strongest deviations: in conditions where the chemical potential varies in the vicinity of the free-carrier band edge. Of course, how the energy dependence of the self-energy would influence the low-density Hall constant is an intriguing open question that is left for future work.

The peak in the temperature-dependent Hall factor could be searched in known 2D materials. With the typical values $\mathfrak{a}=(4~\mathrm{\AA})^2$, $t_x=100$~meV, and $\Gamma=2$~meV, Eq.~(\ref{eq:nc}) gives $n_c=4\times10^{12}$~cm$^{-2}$ in the least favorable isotropic case. Hence the quantum regime is well within the range of carrier concentrations that can be achieved with the field-effect technology. Assuming $n=10^{12}$~cm$^{-2}$, Eq.~(\ref{eq:onsets}) gives $T_x=278$~K and Eq.~(\ref{eq:rmax}) gives $r_{\max}\approx6\times10^5$. The exact values given by the model are $T_x=319$~K and $r_{\max}=1.1\times10^5$. At that temperature, the conductivity is, however, very small: $\sigma_x\sim10^{-4}~\mu$S. Of course, it is not obvious that the predictions of the model should be trusted when applied to real materials, in which the self-energy surely has a complex energy dependence. Further studies are therefore needed, using realistic band structures and self-energies inspired by microscopic models of interacting electrons.

The codes written and the data generated for this study are available at Ref.~\cite{yareta}.

\begin{acknowledgments}
We are thankful to F.\ Wu and A.\ Morpurgo for drawing our attention to the Hall effect at low density. This work was supported by the Swiss National Science Foundation under Division II (Grant No. 2000020-188687). L.R.\ is funded by the Swiss National Science Foundation through Starting Grant No.\ TMSGI2\_211296.
\end{acknowledgments}

\appendix

\section{Current-current correlation function}
\label{app:CRJJ}

In this appendix, we describe the steps leading to Eqs.~(\ref{eq:CRJJ}) and (\ref{eq:Phiab}). We also list a few symmetries of the transport function $\Phi_{\alpha\beta}(E_1,E_2,B)$, that are useful to establish the symmetry properties of the conductivity tensor.

The local current density is given by the functional derivative of the Hamiltonian, Eq.~(\ref{eq:H}), with respect to the vector potential: $j_{\alpha}(\vec{r})=-\delta H/\delta A_{\alpha}(\vec{r})|_{\vec{E}=0}$. The functional derivative is evaluated at $\vec{E}=0$, because the Kubo formula gives the linear response to the electric field. It is therefore convenient to split the vector potential as $\vec{A}=\vec{A}_{\mathrm{el}}+\vec{A}_{\mathrm{mag}}$, where $\vec{A}_{\mathrm{el}}$ is irrotational and vanishes for $\vec{E}=0$, while $\vec{\nabla}\times\vec{A}_{\mathrm{mag}}=\vec{B}$. The interaction $H_{\mathrm{int}}$ being independent of the vector potential, the functional derivative is straightforward and yields the total current,
    \begin{equation}
		J_{\alpha}=-\frac{i|e|}{\hbar}\sum_{\vec{r}_1\vec{r}_2\sigma}
		(\vec{r}_2-\vec{r}_1)_{\alpha}t^0_{\vec{r}_1\vec{r}_2}e^{iA_{\vec{r}_1\vec{r}_2}}
		c^{\dagger}_{\vec{r}_1\sigma}c^{}_{\vec{r}_2\sigma},
    \end{equation}
with $A_{\vec{r}_1\vec{r}_2}=\frac{|e|}{\hbar}\int_{\vec{r}_1}^{\vec{r}_2}d\vec{r}\cdot\vec{A}_{\mathrm{mag}}(\vec{r})$. The retarded correlation function, Eq.~(\ref{eq:CJJ}), is the analytic continuation of the corresponding Matsubara function: $C^R_{J_{\alpha}J_{\beta}}(\omega)=C_{J_{\alpha}J_{\beta}}(i\Omega_n\to\hbar\omega+i0)$ \cite{Mahan-2000}. Here $\Omega_n=2n\pi\kB T$ are even Matsubara frequencies and $C_{J_{\alpha}J_{\beta}}(i\Omega_n)=\int_0^{1/\kB T}d\tau\,e^{i\Omega_n\tau}C_{J_{\alpha}J_{\beta}}(\tau)$ with the imaginary-time function $C_{J_{\alpha}J_{\beta}}(\tau)=-\langle T_{\tau}J_{\alpha}(\tau)J_{\beta}(0)\rangle$ given explicitly by
    \begin{multline}\label{eq:CMJJ}
		C_{J_{\alpha}J_{\beta}}(\tau)=\left(\frac{e}{\hbar}\right)^2
		\sum_{\vec{r}_1\vec{r}_2\sigma}\sum_{\vec{r}_3\vec{r}_4\sigma'}
		(\vec{r}_2-\vec{r}_1)_{\alpha}t^0_{\vec{r}_1\vec{r}_2}
		(\vec{r}_4-\vec{r}_3)_{\beta}t^0_{\vec{r}_3\vec{r}_4}\\
		\times e^{iA_{\vec{r}_1\vec{r}_2}}e^{iA_{\vec{r}_3\vec{r}_4}}
		\langle T_{\tau}c^{\dagger}_{\vec{r}_1\sigma}(\tau)c^{}_{\vec{r}_2\sigma}(\tau)
		c^{\dagger}_{\vec{r}_3\sigma'}(0)c^{}_{\vec{r}_4\sigma'}(0)\rangle.
    \end{multline}
$T_{\tau}$ indicates imaginary-time ordering and $\langle\,\cdots\rangle$ a thermal average with respect to $H$ at $\vec{E}=0$ and finite $\vec{B}$. The assumption of local correlations in a paramagnetic state implies that the thermal average factorizes as
	\begin{multline*}
		\langle T_{\tau}c^{\dagger}_{\vec{r}_1\sigma}(\tau)c^{}_{\vec{r}_2\sigma}(\tau)
		c^{\dagger}_{\vec{r}_3\sigma'}(0)c^{}_{\vec{r}_4\sigma'}(0)\rangle\\
		=\langle T_{\tau}c^{\dagger}_{\vec{r}_1\sigma}(\tau)c^{}_{\vec{r}_4\sigma'}(0)\rangle
		\langle T_{\tau}c^{}_{\vec{r}_2\sigma}(\tau)c^{\dagger}_{\vec{r}_3\sigma'}(0)\rangle\\
		=-\delta_{\sigma\sigma'}G(\vec{r}_4,\vec{r}_1,-\tau)G(\vec{r}_2,\vec{r}_3,\tau),
	\end{multline*}
where $G(\vec{r}_1,\vec{r}_2,\tau)=-\langle T_{\tau}c^{}_{\vec{r}_1\sigma}(\tau)c^{\dagger}_{\vec{r}_2\sigma}(0)\rangle$ is the interacting single-particle Green's function.

The Green's function as defined above isn't gauge invariant and breaks translational symmetry in the presence of a finite (spatially uniform) magnetic field. Indeed, a change of the electromagnetic gauge modifies the Peierls phase according to $A_{\vec{r}_1\vec{r}_2}\to A_{\vec{r}_1\vec{r}_2}+\chi(\vec{r}_2)-\chi(\vec{r}_1)$, where $\chi(\vec{r})$ is the gauge. The gauge invariance of the Hamiltonian requires that the field operators change as $c_{\vec{r}\sigma}\to e^{-i\chi(\vec{r})}c_{\vec{r}\sigma}$. Hence, the Green's function changes as $G(\vec{r}_1,\vec{r}_2,\tau)\to G(\vec{r}_1,\vec{r}_2,\tau)e^{i[\chi(\vec{r}_2)-\chi(\vec{r}_1)]}$. Let us therefore introduce a modified Green's function \cite{Chen-2011, Vucicevic-2021}
	\begin{equation}\label{eq:Gbar}
		\bar{G}(\vec{r}_1-\vec{r}_2,\tau)=e^{-iA_{\vec{r}_1\vec{r}_2}}G(\vec{r}_1,\vec{r}_2,\tau),
	\end{equation}
which is manifestly gauge invariant \footnote{Using the equation-of-motion method for expressing the real-space Green's function, one can verify that $\bar{G}$ is also translation invariant.}. Thanks to its translation invariance in space and time, $\bar{G}$ admits the Fourier representation $\bar{G}(\vec{r},\tau)=\frac{\kB T}{N}\sum_{\vec{k}}\sum_{i\omega_n}\bar{G}(\vec{k},i\omega_n)e^{i(\vec{k}\cdot\vec{r}-\omega_n\tau)}$, where $N$ is the number of $\vec{k}$ points and $\omega_n=(2n+1)\pi\kB T$ are odd Matsubara frequencies. Introducing this in Eq.~(\ref{eq:CMJJ}), we get
     \begin{multline*}
		C_{J_{\alpha}J_{\beta}}(i\Omega_n)=-\left(\frac{e}{\hbar}\right)^2
		\frac{1}{N^2}\sum_{\vec{k}\vec{q}\sigma}\kB T\sum_{i\omega_n}\\
		\times \bar{G}(\vec{k},i\omega_n)\bar{G}(\vec{k}+\vec{q},i\omega_n+i\Omega_n)
		\sum_{\vec{r}_1\vec{r}_2\vec{r}_3\vec{r}_4}
		e^{i\vec{k}\cdot(\vec{r}_4-\vec{r}_1)}
		e^{i(\vec{k}+\vec{q})\cdot(\vec{r}_2-\vec{r}_3)}\\
		\times(\vec{r}_2-\vec{r}_1)_{\alpha}t^0_{\vec{r}_1\vec{r}_2}
		(\vec{r}_4-\vec{r}_3)_{\beta}t^0_{\vec{r}_3\vec{r}_4}
		e^{i(A_{\vec{r}_1\vec{r}_2}+A_{\vec{r}_3\vec{r}_4}+A_{\vec{r}_4\vec{r}_1}+A_{\vec{r}_2\vec{r}_3})}.
    \end{multline*}
This correlation function is gauge invariant, because the quantity $A_{\vec{r}_1\vec{r}_2}+A_{\vec{r}_2\vec{r}_3}+A_{\vec{r}_3\vec{r}_4}+A_{\vec{r}_4\vec{r}_1}$ is proportional to the magnetic flux threading the oriented tetragon defined by the points $(\vec{r}_1,\vec{r}_2,\vec{r}_3,\vec{r}_4)$, and is therefore independent of the gauge:
	\begin{equation}
		A_{\vec{r}_1\vec{r}_2}+A_{\vec{r}_2\vec{r}_3}+A_{\vec{r}_3\vec{r}_4}+A_{\vec{r}_4\vec{r}_1}=
		\frac{|e|}{2\hbar}\vec{B}\cdot(\vec{r}_1-\vec{r}_3)\times(\vec{r}_2-\vec{r}_4).
	\end{equation}
This is readily verified, for instance, in the symmetric gauge $\vec{A}_{\mathrm{mag}}(\vec{r})=\frac{B}{2}(-y,x,0)$, where $A_{\vec{r}_1\vec{r}_2}=\frac{|e|B}{2\hbar}(r_{1x}r_{2y}-r_{1y}r_{2x})=\frac{|e|}{2\hbar}\vec{B}\cdot(\vec{r}_1\times\vec{r}_2)$.

The Green's function admits an integral representation in terms of its spectral function:
    \begin{equation}
		 \bar{G}(\vec{k},i\omega_n)=\int_{-\infty}^{\infty}d\varepsilon\,\frac{\bar{A}(\vec{k},\varepsilon)}
		 {i\omega_n+\mu-\varepsilon}.
    \end{equation}
$\bar{A}(\vec{k},\varepsilon)$ contains the fractal Hofstadter physics and cannot be expanded in powers of $B$. However, because the Hofstadter spectrum is even in $B$, one can conclude that $\bar{A}(\vec{k},\varepsilon)$ formally deviates from its value at $B=0$ by corrections of order $B^2$. Up to first order in $B$, it is therefore sufficient to replace $\bar{A}(\vec{k},\varepsilon)$ by its zero-field value, which is $A(E_{\vec{k}},\varepsilon)$ given in Eq.~(\ref{eq:A}). After inserting the spectral representation of $\bar{G}$, standard techniques allow one to perform the sum over odd Matsubara frequencies \cite{Mahan-2000}. A further shift of $\vec{k}$ by $-\vec{q}/2$ to achieve a more symmetric form and introduction of the relative vectors $\vec{\rho}_1=\vec{r}_2-\vec{r}_1$, $\vec{\rho}_2=\vec{r}_4-\vec{r}_3$, and $\vec{r}=\vec{r}_1-\vec{r}_3$ directly leads to Eqs.~(\ref{eq:CRJJ}) and (\ref{eq:Phiab}).

The transport function defined in Eq.~(\ref{eq:Phiab}) obeys the following symmetries:
	\begin{subequations}\begin{align}
		\label{eq:sym1}
		\Phi_{\alpha\beta}(E_1,E_2,B)&=\Phi^*_{\alpha\beta}(E_1,E_2,-B)\\
		\label{eq:sym2}
		\Phi_{\alpha\beta}(E_1,E_2,B)&=\Phi^*_{\alpha\beta}(E_2,E_1,B)\\
		\label{eq:sym3}
		\Phi_{\alpha\beta}(E_1,E_2,B)&=\Phi_{\beta\alpha}(E_2,E_1,B).
	\end{align}\end{subequations}
The first may be checked by changing $(\vec{r},\vec{\rho}_1,\vec{\rho}_2)$ into $(-\vec{r},-\vec{\rho}_1,-\vec{\rho}_2)$; the second by changing $(\vec{q},\vec{\rho}_1,\vec{\rho}_2,\vec{r})$ into $(-\vec{q},-\vec{\rho}_1,-\vec{\rho}_2,\vec{r}+\vec{\rho}_1-\vec{\rho}_2)$; and the third by changing $(\vec{q},\vec{\rho}_1,\vec{\rho}_2,\vec{r})$ into $(-\vec{q},\vec{\rho}_2,\vec{\rho}_1,-\vec{r})$. Using these symmetries, one verifies that $\sigma_{\alpha\beta}(\omega,B)=\sigma^*_{\beta\alpha}(-\omega,-B)$. In the dc limit, this gives the Onsager-Casimir relation $\sigma_{\alpha\beta}(B)=\sigma_{\beta\alpha}(-B)$ \cite{Onsager-1931, Casimir-1945}. Equation~(\ref{eq:sym1}) implies that $\mathrm{Re}\,\Phi_{\alpha\beta}(E_1,E_2,B)$ and $\mathrm{Im}\,\Phi_{\alpha\beta}(E_1,E_2,B)$ are even and odd functions of $B$, respectively, which fixes the parity of Eqs.~(\ref{eq:sigmae}) and (\ref{eq:sigmao}) in the field. Equation~(\ref{eq:sym3}) furthermore implies that $\sigma_{\beta\alpha}^{(e)}(B)=\sigma_{\alpha\beta}^{(e)}(B)$ and $\sigma_{\beta\alpha}^{(o)}(B)=-\sigma_{\alpha\beta}^{(o)}(B)$. The latter shows that the part odd in the field is anti diagonal, $\sigma_{\alpha\alpha}^{(o)}(B)=0$. Making use of $\sigma_{xy}=-\sigma_{yx}$, the former shows that the even part is diagonal: $\sigma_{\alpha\beta}^{(e)}(B)\propto\delta_{\alpha\beta}$.

\section{Weak-field expansion of the dc conductivity}
\label{app:expansion}

The magnetic field dependence of the dc conductivity defined by Eqs.~(\ref{eq:DC})--(\ref{eq:Phiab}) stems from the quantity
	\begin{equation}
		\Delta(\vec{q},\vec{\rho},B)=\frac{1}{N}\sum_{\vec{r}}e^{i\vec{q}\cdot\vec{r}}
		e^{-\frac{i|e|}{2\hbar}B\hat{\vec{z}}\cdot\vec{\rho}\times\vec{r}}.
	\end{equation}
Its zero-field limit gives a Kronecker symbol at $\vec{q}=\vec{0}$:
	\begin{equation}
		\Delta(\vec{q},\vec{\rho},0)=\frac{1}{N}\sum_{\vec{r}}e^{i\vec{q}\cdot\vec{r}}=\delta_{\vec{q}\vec{0}}.
	\end{equation}
The transport function at this order is, therefore,
	\begin{multline*}
		\Phi_{\alpha\beta}(E_1,E_2,0)=-\left(\frac{e}{\hbar}\right)^2\frac{1}{V}\sum_{\vec{k}\sigma}
		\delta\left(E_1-E_{\vec{k}}\right)\delta\left(E_2-E_{\vec{k}}\right)\\
		\times\sum_{\vec{\rho}_1\vec{\rho}_2}
		\rho_{1\alpha}t^0_{\vec{\rho}_1}e^{i\vec{k}\cdot\vec{\rho}_1}
		\rho_{2\beta}t^0_{\vec{\rho}_2}e^{i\vec{k}\cdot\vec{\rho}_2}.
	\end{multline*}
Introducing the group velocity $v_{\vec{k}\alpha}=(1/\hbar)\partial E_{\vec{k}}/\partial k_{\alpha}$, we can substitute $\sum_{\vec{\rho}}\rho_{\alpha}t^0_{\vec{\rho}}e^{i\vec{k}\cdot\vec{\rho}}$ by $-i\hbar v_{\vec{k}\alpha}$. Since we only consider systems with the inversion symmetry $E_{\vec{k}}=E_{-\vec{k}}$, the components of the group velocity are odd functions of $\vec{k}$. Hence, $\Phi_{\alpha\beta}$ vanishes by symmetry if $\alpha\neq\beta$ and we have
	\begin{equation*}
		\Phi_{\alpha\beta}(E_1,E_2,0)=\delta_{\alpha\beta}\delta(E_1-E_2)e^2\frac{1}{V}\sum_{\vec{k}\sigma}
		v_{\vec{k}\alpha}^2\delta\left(E_1-E_{\vec{k}}\right),
	\end{equation*}
which corresponds to Eq.~(\ref{eq:Phi0}).

At first order in $B$, we have
	\begin{equation}
		\left.\frac{d}{dB}\Delta(\vec{q},\vec{\rho},B)\right|_{B=0}
		=-\frac{|e|}{2\hbar}\hat{\vec{z}}\cdot\vec{\rho}\times
		\vec{\nabla}_{\vec{q}}\frac{1}{N}\sum_{\vec{r}}e^{i\vec{q}\cdot\vec{r}}.
	\end{equation}
This formally involves the gradient of the Kronecker symbol $\delta_{\vec{q}\vec{0}}$, which is only well-defined after taking the continuum limit for the momenta. Implying this limit, we evaluate the $\vec{q}$ sum in Eq.~(\ref{eq:Phiab}) by parts as $\sum_{\vec{q}}F(\vec{q})\vec{\nabla}_{\vec{q}}\delta_{\vec{q}\vec{0}}=-\sum_{\vec{q}}\delta_{\vec{q}\vec{0}}\vec{\nabla}_{\vec{q}}F(\vec{q})=-\vec{\nabla}_{\vec{q}}F(\vec{q})|_{\vec{q}=\vec{0}}$. Upon calculating $d\Phi_{\alpha\beta}(E_1,E_2,B)/dB$ at $B=0$, the $\vec{\nabla}_{\vec{q}}$ yields one term from acting on the phase factors and two terms from acting on the Dirac delta functions. The former vanishes, as it is proportional to $(\vec{\rho}_1-\vec{\rho}_2)\times(\vec{\rho}_1-\vec{\rho}_2)$. We get
	\begin{multline}\label{eq:dPhiab}
		\left.\frac{d}{dB}\Phi_{\alpha\beta}(E_1,E_2,B)\right|_{B=0}=
		\frac{i|e|^3\hbar}{4}\frac{1}{V}\sum_{\vec{k}\sigma}\\
		\times\left[\delta'\left(E_1-E_{\vec{k}}\right)\delta\left(E_2-E_{\vec{k}}\right)
		-\delta\left(E_1-E_{\vec{k}}\right)\delta'\left(E_2-E_{\vec{k}}\right)\right]\\
		\times\left(
		\frac{v_{\vec{k}x}v_{\vec{k}\beta}}{m_{\vec{k}y\alpha}}
		-\frac{v_{\vec{k}y}v_{\vec{k}\beta}}{m_{\vec{k}x\alpha}}
		-\frac{v_{\vec{k}\alpha}v_{\vec{k}x}}{m_{\vec{k}y\beta}}
		+\frac{v_{\vec{k}\alpha}v_{\vec{k}y}}{m_{\vec{k}x\beta}}
		\right),
	\end{multline}
where $\delta'$ stands for the derivative of the Dirac delta and the mass tensor $1/m_{\vec{k}\alpha\beta}=(1/\hbar^2)\partial^2E_{\vec{k}}/(\partial k_{\alpha}\partial k_{\beta})$ has been introduced. This result is purely imaginary and thus only contributes to $\sigma_{\alpha\beta}^{(o)}(B)$, as expected. It is also odd under the exchange of $\alpha$ and $\beta$ and vanishes as it should if $\alpha=\beta$.

When Eq.~(\ref{eq:dPhiab}) is inserted in Eq.~(\ref{eq:sigmao}), the resulting $E_1$ or $E_2$ integral involving $\delta'$ can be evaluated by parts using, for instance,
	$
		\int_{-\infty}^{\infty}dE_1\,\delta'(E_1-E_{\vec{k}})A(E_1,\varepsilon)
		=-\int_{-\infty}^{\infty}dE_1\,\delta(E_1-E_{\vec{k}})A'(E_1,\varepsilon),
	$
with $A'(E,\varepsilon)$ denoting the derivative of the spectral function with respect to the first argument. Upon performing this substitution, we encounter the integral
	\begin{align*}
		I&=\int_{-\infty}^{\infty}d\varepsilon_1d\varepsilon_2\,
		\frac{f(\varepsilon_1-\mu)-f(\varepsilon_2-\mu)}{(\varepsilon_1-\varepsilon_2)^2}\\
		&\quad\times
		\left[A(E,\varepsilon_1)A'(E,\varepsilon_2)-A'(E,\varepsilon_1)A(E,\varepsilon_2)\right].
	\end{align*}
Remarkably, if the spectral function is causal, this integral can be recast as
	\begin{equation}\label{eq:off-on}
		I=\frac{4\pi^2}{3}\int_{-\infty}^{\infty}d\varepsilon\,[-f'(\varepsilon-\mu)]A^3(E,\varepsilon).
	\end{equation}
This leads to Eqs.~(\ref{eq:sigma1}) and (\ref{eq:Phi1}). To prove Eq.~(\ref{eq:off-on}), we rewrite
	\begin{multline*}
		I=2\int_{-\infty}^{\infty}d\varepsilon_1\,f(\varepsilon_1-\mu)A(E,\varepsilon_1)
		\int_{\infty}^{\infty}d\varepsilon_2\,\frac{A'(E,\varepsilon_2)}{(\varepsilon_1-\varepsilon_2)^2}\\
		-2\int_{-\infty}^{\infty}d\varepsilon_1\,f(\varepsilon_1-\mu)A'(E,\varepsilon_1)
		\int_{\infty}^{\infty}d\varepsilon_2\,\frac{A(E,\varepsilon_2)}{(\varepsilon_1-\varepsilon_2)^2},
	\end{multline*}
where a principal value is understood in the $\varepsilon_2$ integrals. We now use the causality of the self-energy, which implies (and requires)
	\begin{align*}
		\mathrm{Re}\,G(E,\varepsilon_1)&=\int_{\infty}^{\infty}d\varepsilon_2\,
		\frac{A(E,\varepsilon_2)}{\varepsilon_1-\varepsilon_2}\\
		\frac{d}{d\varepsilon_1}\mathrm{Re}\,G(E,\varepsilon_1)&=
		-\int_{\infty}^{\infty}d\varepsilon_2\,
		\frac{A(E,\varepsilon_2)}{(\varepsilon_1-\varepsilon_2)^2},
	\end{align*}
with $G(E,\varepsilon)$ the retarded Green's function. It follows that
	\begin{multline*}
		I=-2\int_{-\infty}^{\infty}d\varepsilon\,f(\varepsilon-\mu)\\
		\times\underbrace{\left[A(E,\varepsilon)
		\frac{d}{d\varepsilon}\mathrm{Re}\,G'(E,\varepsilon)-A'(E,\varepsilon)
		\frac{d}{d\varepsilon}\mathrm{Re}\,G(E,\varepsilon)\right]}_{\displaystyle
		=-\frac{2\pi^2}{3}\frac{d}{d\varepsilon}A^3(E,\varepsilon).}
	\end{multline*}
We have used Eq.~(\ref{eq:A}) and the explicit form of the Green's function, $G(E,\varepsilon)=1/[\varepsilon-E-\Sigma(\varepsilon)]$. An integration by parts then gives Eq.~(\ref{eq:off-on}).

\section{Relation with the semiclassical theory}
\label{app:Boltzmann}

The Boltzmann expressions in the isotropic relaxation-time approximation may be derived from Eqs.~(\ref{eq:sigma0})--(\ref{eq:Phi1}) under the assumptions that (i) the spectral function is narrower than the derivative of the Fermi distribution, such that $-f'(\varepsilon-\mu)$ may be replaced by $-f'(E-\mu)$ and (ii) the self-energy takes the form $\Sigma(\varepsilon)=-i\hbar/(2\tau)$, where $\tau$ is the relaxation time. Using the identities $\int_{-\infty}^{\infty}d\varepsilon\,A^2(E,\varepsilon)=1/(2\pi\Gamma)$ and $\int_{-\infty}^{\infty}d\varepsilon\,A^3(E,\varepsilon)=(3/2)/(2\pi\Gamma)^2$, one thus arrives at the approximations
	\begin{align*}
		\sigma_{\alpha}^{(0)}&\approx e^2\sum_{\vec{k}\sigma}[-f'(E_{\vec{k}}-\mu)]\tau
		v_{\vec{k}\alpha}^2\\
		\sigma_{xy}^{(1)}&\approx-|e|^3\sum_{\vec{k}\sigma}[-f'(E_{\vec{k}}-\mu)]\tau^2\\
		&\quad\times
		\left[\frac{1}{2}\left(\frac{v_{\vec{k}x}^2}{m_{\vec{k}yy}}+\frac{v_{\vec{k}y}^2}{m_{\vec{k}xx}}\right)
		-\frac{v_{\vec{k}x}v_{\vec{k}y}}{m_{\vec{k}xy}}\right],
	\end{align*}
which, apart from the $(x,y)$ symmetrization at the last line, coincide with the semiclassical formula given in the literature \cite{Ziman-1960, Fukuyama-1969-2, Ong-1991}.

\section{\boldmath Exact integrals for the case $\Sigma(\varepsilon)=-i\Gamma$}
\label{app:convolutions}

The self-energy $\Sigma(\varepsilon)=-i\Gamma$ obeys causality, because the Kramers--Kronig transform of a constant is zero. Consistently, the spectral function $A(E,\varepsilon)=\frac{\Gamma/\pi}{(\varepsilon-E)^2+\Gamma^2}$ is properly normalized to $\int_{-\infty}^{\infty}d\varepsilon\,A(E,\varepsilon)=1$. The integrals that involve the Fermi distribution and the spectral function have a closed form in terms of polygamma functions $\psi_n(z)$, where $z=\frac{1}{2}-i\frac{E-\mu+i\Gamma}{2\pi\kB T}$. The $\varepsilon$ integral in Eq.~(\ref{eq:n}) reads
	\begin{align*}
		\int_{-\infty}^{\infty}d\varepsilon\,f(\varepsilon-\mu)A(E,\varepsilon)&=\frac{1}{2}+\frac{1}{\pi}
		\mathrm{Im}\,\psi_0(z)\\
		&=\frac{1}{2}-\frac{1}{\pi}\tan^{-1}\left(\frac{E-\mu}{\Gamma}\right)\quad(T=0).
	\end{align*}
The integrals appearing in Eqs.~(\ref{eq:sigma0}) and (\ref{eq:sigma1}) are trivial at $T=0$ and otherwise read
	\begin{align*}
		&\int_{-\infty}^{\infty}d\varepsilon\,[-f'(\varepsilon-\mu)]A^2(E,\varepsilon)\\
		&\qquad=\frac{\zeta}{2(\pi\Gamma)^2}\mathrm{Re}\left[\psi_1(z)-\zeta\psi_2(z)\right]\\
		&\int_{-\infty}^{\infty}d\varepsilon\,[-f'(\varepsilon-\mu)]A^3(E,\varepsilon)\\
		&\qquad=\frac{3\zeta}{8(\pi\Gamma)^3}\mathrm{Re}\left[
        \psi_1(z)-\zeta\psi_2(z)+\frac{\zeta^2}{3}\psi_3(z)\right],
	\end{align*}
where $\zeta=\frac{\Gamma}{2\pi\kB T}$.

\section{Exact asymptotic results and approximations}
\label{app:asymptotics}

\subsection{\boldmath $T=0$}
\label{app:asymptoticsT=0}

At zero temperature, the transport equations are
	\begin{align*}
		n&=\int_{-\infty}^{\infty}dE\,N_0(E)
		\left[\frac{1}{2}-\frac{1}{\pi}\tan^{-1}\left(\frac{E-\mu}{\Gamma}\right)\right]\\
		\sigma_{\alpha}^{(0)}&=\pi\hbar\int_{-\infty}^{\infty}dE\,\Phi_{\alpha}^{(0)}(E)A^2(E,\mu)\\
		\sigma_{xy}^{(1)}&=\hbar\int_{-\infty}^{\infty}dE\,\Phi^{(1)}(E)A^3(E,\mu).
	\end{align*}
In the semiclassical regime, where $\Gamma$ is the smallest energy scale, the transport functions may be expanded around $E=\mu$. The leading terms give
	\begin{align*}
		\sigma_{\alpha}^{(0)}&=\pi\hbar\Phi_{\alpha}^{(0)}(\mu)\int_{-\infty}^{\infty}dE\,A^2(E,\mu)
		=\frac{\hbar}{2\Gamma}\Phi_{\alpha}^{(0)}(\mu)\\
		\sigma_{xy}^{(1)}&=\hbar\Phi^{(1)}(\mu)\int_{-\infty}^{\infty}dE\,A^3(E,\mu)
		=\frac{3\hbar}{8\pi^2\Gamma^2}\Phi^{(1)}(\mu).
	\end{align*}
This leads to Eq.~(\ref{eq:RH1}) of the main text. To estimate the crossover density in Fig.~\ref{fig:RH-T=0}, we fix $\mu=-D$ in the equation giving the density. Since the term in square brackets drops rapidly for energies away from the band edge, we can use the approximation
	\begin{align*}
		n_c&\approx N_0(-D)\int_{-D}^DdE\,
		\left[\frac{1}{2}-\frac{1}{\pi}\tan^{-1}\left(\frac{E+D}{\Gamma}\right)\right]\\
		&=\frac{\Gamma}{2\pi^2\mathfrak{a}\sqrt{|t_xt_y|}}\left[1+\ln\left(\frac{2D}{\Gamma}\right)
		+\mathscr{O}\left(\frac{\Gamma^2}{D^2}\right)\right].
	\end{align*}
In the quantum regime, where $n$ approaches zero at finite $\Gamma$, we can expand the equations for $\mu\to-\infty$:
	\begin{align*}
		n&=\int_{-\infty}^{\infty}dE\,N_0(E)
		\left[-\frac{\Gamma}{\pi\mu}-\frac{\Gamma E}{\pi\mu^2}+\mathscr{O}\left(\frac{1}{\mu^3}\right)\right]\\
		\sigma_{\alpha}^{(0)}&=\pi\hbar\int_{-\infty}^{\infty}dE\,\Phi_{\alpha}^{(0)}(E)
		\left[\frac{\Gamma^2}{\pi^2\mu^4}+\frac{4\Gamma^2E}{\pi^2\mu^5}
		+\mathscr{O}\left(\frac{1}{\mu^6}\right)\right]\\
		\sigma_{xy}^{(1)}&=\hbar\int_{-\infty}^{\infty}dE\,\Phi^{(1)}(E)
		\left[\frac{\Gamma^3}{\pi^3\mu^6}+\frac{6\Gamma^3E}{\pi^3\mu^7}
		+\mathscr{O}\left(\frac{1}{\mu^8},\frac{E^2}{\mu^8}\right)\right].
	\end{align*}
Considering the symmetry of the DOS and transport functions, the relevant moments are
	\begin{align*}
		\int_{-\infty}^{\infty}dE\,N_0(E)&=\frac{2}{\mathfrak{a}}\\
		\int_{-\infty}^{\infty}dE\,\Phi_{\alpha}^{(0)}(E)
		&=\left(\frac{e}{\hbar}\right)^2\frac{4a_{\alpha}^2t_{\alpha}^2}{\mathfrak{a}}\\
		\int_{-\infty}^{\infty}dE\,E\,\Phi^{(1)}(E)
		&=\left(\frac{|e|}{\hbar}\right)^3\frac{16\pi^2}{3}\mathfrak{a}t_x^2t_y^2,
	\end{align*}
where $a_x$ and $a_y$ are the lattice parameters and $\mathfrak{a}=a_xa_y$. Hence, we arrive at
	\begin{align*}
		n&=-\frac{1}{\mathfrak{a}}\frac{2\Gamma}{\pi\mu}+\mathscr{O}(1/\mu^3)\\
		\sigma_{\alpha}^{(0)}&=\frac{e^2a_{\alpha}^2t_{\alpha}^2}{\hbar\mathfrak{a}}\frac{4\Gamma^2}{\pi\mu^4}
		+\mathscr{O}(1/\mu^6)\\
		\sigma_{xy}^{(1)}&=\frac{|e|^3\mathfrak{a}t_x^2t_y^2}{\hbar^2}\frac{32\Gamma^3}{\pi\mu^7}
		+\mathscr{O}(1/\mu^9),
	\end{align*}
which leads to Eq.~(\ref{eq:RH4}).

\subsection{\boldmath $T>0$}
\label{app:asymptoticsT>0}

At low temperature, we rely on a Sommerfeld expansion. For the density in Eq.~(\ref{eq:n}), we integrate by parts to let the quantity $-f'(\varepsilon-\mu)$ appear. We change variables from $\varepsilon$ to $u=(\varepsilon-\mu)/(\kB T)$, expand the integrand in powers of $T$, and perform the integrals using $\int_{-\infty}^{\infty}du\,\frac{e^u}{(e^u+1)^2}\{1,u,u^2\}=\{1,0,\pi^2/3\}$. The condition $n(T)=n(0)$ allows us to deduce $\mu'(0)=0$, where $\mu'(0)$ is the first derivative of the chemical potential with respect to $\kB T$. A closed expression for the second derivative can also be deduced in the low-density regime, after sending $\mu_0\equiv\mu(T=0)$ to $-\infty$ and using the relation $\mu_0=-2\Gamma/(\pi n\mathfrak{a})$ derived above. This expression is $\mu''(0)=-\pi^3n\mathfrak{a}/(3\Gamma)$. Proceeding similarly with Eqs.~(\ref{eq:sigma0}) and (\ref{eq:sigma1}), we arrive at the relations
	\begin{align*}
		\frac{\sigma_{\alpha}^{(0)}(T)}{\sigma_{\alpha}^{(0)}(0)}&=
		1+\frac{\pi^4}{2\Gamma^2}(n\mathfrak{a})^2(\kB T)^2\\
		\frac{\sigma^{(1)}_{xy}(T)}{\sigma^{(1)}_{xy}(0)}&=
		1+\frac{7\pi^4}{4\Gamma^2}(n\mathfrak{a})^2(\kB T)^2\\
		\frac{r(T)}{r(0)}&=1+\frac{3\pi^4}{4\Gamma^2}(n\mathfrak{a})^2(\kB T)^2,
	\end{align*}
which are valid for $T\to0$ and $n\to0$ at finite $\Gamma$. At higher density, when $\mu_0$ lies inside the band, instead of the expansion around $\mu_0=-\infty$ we expand the transport functions around $E=\mu_0$. Thus we find $\mu''(0)=-(\pi^2/3)N_0'(\mu_0)/N_0(\mu_0)$ and, for the conductivities,
	\begin{align*}
		\frac{\sigma_{\alpha}^{(0)}(T)}{\sigma_{\alpha}^{(0)}(0)}&=
		1-\frac{\pi^2}{6}\left[\frac{N_0'(\mu_0)}{N_0(\mu_0)}
		\frac{{\Phi_{\alpha}^{(0)}}'(\mu_0)}{\Phi_{\alpha}^{(0)}(\mu_0)}
		-\frac{{\Phi_{\alpha}^{(0)}}''(\mu_0)}{\Phi_{\alpha}^{(0)}(\mu_0)}\right](\kB T)^2\\
		\frac{\sigma^{(1)}_{xy}(T)}{\sigma^{(1)}_{xy}(0)}&=
		1-\frac{\pi^2}{6}\left[\frac{N_0'(\mu_0)}{N_0(\mu_0)}
		\frac{{\Phi^{(1)}}'(\mu_0)}{\Phi^{(1)}(\mu_0)}
		-\frac{{\Phi^{(1)}}''(\mu_0)}{\Phi^{(1)}(\mu_0)}\right](\kB T)^2.
	\end{align*}
We estimate the onset temperatures $T_x$, $T_y$, and $T_{xy}$ in Fig.~\ref{fig:T2} as follows. In the regime $\Gamma<\kB T\ll D$, the spectral function is narrower than the derivative of the Fermi function. We therefore replace in Eqs.~(\ref{eq:sigma0}) and (\ref{eq:sigma1}) $f'(\varepsilon-\mu)$ by $f'(E-\mu)$, such that the $\varepsilon$ integrals become trivial. With the low-density regime in mind, we further note that for $\kB T\ll D$, the function $f'(E-\mu)$ is sharp compared with the transport function and peaked near the band bottom. We therefore replace the transport functions by their linearizations close to the band bottom, which are indicated in the legend of Fig.~\ref{fig:N0-Phi}. We thus arrive at Eq.~(\ref{eq:sigma-above-onset}).

We close this appendix by showing that, in the high-temperature regime $\Gamma\ll\kB T<D$, the Hall constant recovers approximately the classical value Eq.~(\ref{eq:RH1}), irrespective of the density. In that regime, we can replace $f'(\varepsilon-\mu)$ by $f'(E-\mu)$, because the derivative of the Fermi function is broad compared with the spectral function, and we can replace the transport function by its value at $E=\mu$, because the derivative of the Fermi function is still much sharper than the transport function. This directly leads to the result
\begin{equation*}
	\RH\approx\frac{3}{2\pi^2\hbar}\frac{\Phi^{(1)}(\mu)}{\Phi^{(0)}_x(\mu)\Phi^{(0)}_y(\mu)}
	\quad(\Gamma\ll\kB T<D),
\end{equation*}
which is the same as the zero-temperature Hall constant for $\Gamma\to0$.

\end{document}